# Time-resolved two-photon interference of weak coherent pulses

Heonoh Kim,[1] Osung Kwon,[2,a)] and Han Seb Moon,[1,a)]

[1]Department of Physics, Pusan National University, Geumjeong-Gu, Busan 46241, Korea
[2]Affiliated Institute of Electronics and Telecommunications Research Institute, Daejeon 34044, South Korea

[a)]Authors to whom correspondence should be addressed: hsmoon@pusan.ac.kr and oskwon@nsr.re.kr

The observation of the Hong-Ou-Mandel (HOM)-type two-photon interference (TPI) has played an important role in the development of photonic quantum technologies. The time-resolved coincidence-detection technique has been effectively used to identify and characterize the TPI phenomena of long-coherence optical fields. Here, we report on the experimental demonstration of the TPI of two phase-randomized weak coherent pulses with time-resolved coincidence detection. The mutual coherence time between the two weak coherent lights is determined by applying a frequency noise to one of the two interfering lights. We analyze the HOM-type TPI-fringe visibility according to the ratio of the coherence time to the pulse duration.

The interference of light from classical light or single-photon sources lies at the heart of photonic quantum information technologies [1-4]. In particular, the observation of the two-photon interference (TPI) of two independent phase-randomized weak coherent pulses prepared in a single-photon level has played a key role in the practical realization of quantum communication protocols like measurement-device-independent quantum key distribution (MDI-QKD) [5-12]. The most important aspect in the successful implementation of the MDI-QKD protocol is the observation of a high-visibility Hong-Ou-Mandel (HOM) fringe [13]. The TPI visibility can be strongly affected by intrinsic distinguishability factors such as the frequency and polarization between the two interfering photons, which can be quantified by the degree of mutual coherence between the photons.

The usual approach to observe the HOM interference fringe is to overlap two indistinguishable photons at a beam splitter within their coherence time, which is significantly shorter than the timing resolution of the employed single-photon detectors (SPDs). Therefore, the characteristic HOM fringe shape is observed as a coincidence dip corresponding to the function of the arrival-time difference or relative optical delay between the two photons at the beam splitter [12,13].

However, on employing a light source with a very long coherence time relative to the time resolution of the SPDs, it is very difficult to implement the optical-delay based observation of the TPI full-fringe shape, whereas the time-resolved two-photon detection technique can be effectively utilized to observe and characterize the TPI effect [14,15]. Particularly, we note that the time-resolved coincidence-counting approach was initially used to analyze the temporal distribution of pseudo-thermal light source [16]. Recently, the time-resolved coincidence measurement has been employed to characterize the TPI of long-coherence continuous-mode weak coherent lights [17,18].

To date, the time-resolved measurement of the TPI effect of two phase-independent pulse-mode photons has been commonly utilized to verify indistinguishability between single photons generated from quantum systems such as the atom-cavity system [15], quantum dots [19], and nitrogen-vacancy centers [20], organic dye molecule [21], as well as from two highly different light sources at the single-photon level [22].

When the two photons are overlapped temporally within the pulse duration, the TPI effect can be identified as a reduction in the integrated coincidence counting events [15,19-22]. Therefore, the characteristic interference fringe reveals different shape upon varying the mutual coherence time between the two interfering photons. The time-resolved TPI measurement can also be a useful tool to analyze the effects of mutual coherence between two interfering weak coherent lights with very long coherence times on the TPI fringe visibility [8].

In this paper, we apply the time-resolved two-photon detection technique to observe the characteristic TPI fringe shape of phase-randomized pulse-mode weak coherent lights. The mutual coherence time of the two interfering weak coherent lights is determined by observing the HOM fringe when a frequency noise is applied to one of the two interfering lights. The characteristic TPI fringe patterns depending on the mutual coherence time are observed within a fixed pulse duration. We investigate the TPI fringe visibility observed upon varying the ratio of the mutual coherence time to the pulse duration. Furthermore, we show that the conventional HOM-dip fringe shape can be obtained from the measured time-resolved TPI fringe by bounding the coincidence time window. This approach provides a comprehensive understanding of the TPI fringe shape measured using time-resolved two-photon detection via its comparison with the conventional HOM-dip fringe for photons with very short coherence times.

In the case where pulse-mode phase-randomized weak coherent lights are employed to observe the TPI effect based on time-resolved coincidence counting, the shape of the cross-correlation histogram obtained by coincidence counting with two SPDs is given by the convolution of the two input pulse shapes. Therefore, for two identical square-shaped pulses with orthogonal polarizations, the time-dependent cross-correlation function affords the triangular-shaped histogram as per the following equation:.

$$P(t;\perp) = 1 - \frac{|t|}{T_p}, \quad (1)$$

where $t$ denotes the detection-time difference between the two input photons at the SPDs and $T_p$ is the full-width at half maximum (FWHM) of the pulse duration. Eq. (1) corresponds to the case where the TPI does not occur, and thus, it represents the convolved input pulse shape.

On the other hand, for the two input pulses with parallel polarizations, the cross-correlation function can be expressed as

$$P(t;\|) = \left(1 - \frac{|t|}{T_p}\right) \cdot \left[1 - V_m \exp\left(-\frac{t^2}{T_c^2}\right)\right], \quad (2)$$

where $T_c$ denotes the half-width at 1/e maximum of the mutual coherence time of the two input pulses and $V_m$ is the maximum visibility for $T_c \gg T_p$. Under ideal conditions, $V_m$ is estimated as 0.5 for weak coherent lights [23,24]. Now, Eq. (2) corresponds to the case where TPI occurs, and thus, it represents the characteristic TPI fringe shape obtained with phase-randomized pulse-mode weak coherent lights. Here, we assume the Gaussian-shaped mutual coherence function. The first term on the right-hand side of Eq. (2) corresponds to the convolved pulse shape in Eq. (1) thereby restricting the temporal range of coincidence counting; on the other hand, the second term represents the TPI fringe revealed within the pulse duration.

It is noteworthy that Eq. (2) has the same form as that of the HOM-type TPI fringe measured as a function of the arrival-time difference $\delta t$: $N_c(\delta t) = N_\infty \left[1 - V \exp(-\sigma^2 \delta t^2)\right]$, where $N_c$ represents the measured coincidence, $N_\infty$ the coincidences for $\delta t \gg T_c$, $V$ the TPI-fringe visibility, and $\sigma$ the Gaussian-shaped spectral bandwidth. Consequently, the first term in Eq. (2) is analogous to the total coincidences within the resolving time ($T_R$) for continuous-mode coherent lights with a very short coherence time ($T_c \ll T_R$).

The TPI-fringe visibility is defined using the two integrated coincidence counting distributions from Eqs. (1) and (2) as follows:

$$V = 1 - \frac{\int_{-T_p}^{+T_p} P(t;\|) dt}{\int_{-T_p}^{+T_p} P(t;\perp) dt}. \quad (3)$$

Consequently, the TPI-fringe visibility can be expressed as

$$V = 1 - V_m \left[\mu^2 - \exp\left(-\frac{1}{\mu^2}\right)\mu^2 + 2 - \mu\sqrt{\pi}\mathrm{erf}\left(\frac{1}{\mu}\right)\right], \quad (4)$$

where $\mu$ denotes the ratio of the coherence time to the pulse duration ($T_c/T_p$). From Eq. (4) we can estimate the upper bound to obtain the TPI-fringe visibility according to the value of $T_c/T_p$ (or $T_p/T_c$). Interestingly, if we employ Gaussian-shaped input pulses, the TPI-fringe visibility is given by $V_m\sqrt{\mu^2/(1+\mu^2)}$; however, the visibility exhibits a tendency very similar to that expressed by Eq. (4).

Figure 1 shows the schematic of our experiment to observe the TPI of the two phase-randomized weak coherent pulses. Narrow-linewidth continuous-wave (CW) mode coherent light is emitted from a wavelength-tunable diode laser (New Focus, Velocity TLB-6312, frequency of <300 kHz and wavelength tuning range of 776 ~ 781 nm), which is strongly attenuated to the single-photon level by an optical attenuator consisting of multiple stacked neutral density filters. The resulting linearly polarized weak coherent light is then coupled to a polarization-maintaining single-mode fiber for matching the polarization orientation with the fiber-based modulators and polarization-based Mach-Zehnder interferometer. In the experiment, the intensity modulator (IM: EOSPACE) is used only for pulse-mode operation, which is performed at the input port of the interferometer to prepare the identical pulses contributing to the TPI.

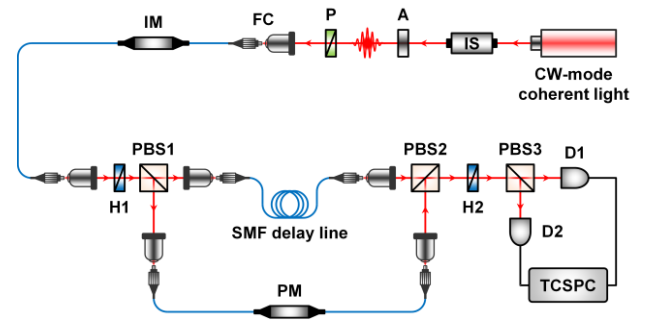

Fig. 1. Experimental setup to observe the two-photon interference of two phase-randomized weak coherent pulses. IS: optical isolator, A: attenuator, P: linear polarizer, FC: single-mode fiber coupler, IM: intensity modulator, H: half-wave plate, PBS: polarizing beam-splitter, PM: phase modulator, D: single-photon detector, TCSPC: time-correlated single-photon counter.

For the HOM two-photon interferometer employing two phase-randomized weak coherent lights, we use a polarization-based interferometer that consists of two half-wave plates (H1 and H2) and three polarizing beam-splitters (PBS1, PBS2, and PBS3) as shown in Fig. 1. One half-wave plate H1 (with its axis oriented at 22.5°) is used to divide equally the input polarization into the two paths of the interferometer by PBS1. Meanwhile another half-wave plate H2 plays a role to erase the polarization information in a polarization-based interferometer. A phase modulator (PM: EOSPACE) is employed to introduce a time-dependent frequency difference between the two weak coherent lights traversing the two interferometer arms caused by adding variable frequency noise. The frequency noise is quantified by the standard deviation of the signal frequency applied to the PM, which is generated by randomly changing the voltage signals over the range from 0 to 5 V using a function generator. The mutual coherence time of the two interfering weak coherent lights is determined by the bandwidth of the function generator, which is limited to ~20 MHz in our experiment; therefore, in the experiment the mutual coherence time could be varied down to ~50 ns. An additional phase-randomization mechanism based on a piezo-electric transducer is used to eliminate the relative phase relation between the two interferometer arms (not shown in Fig. 1).

The single-mode-fiber delay line in one of the paths is used to match the two path lengths in the two interferometer arms. It is noted here that the use of the delay line is not necessary in the continuous-mode operation and time-resolved measurement of the TPI fringe. The two SPDs (SPCM-AQRH, Excelitas), D1 and D2, are positioned at the two output ports of PBS3 and the output signals are fed to the TCSPC (PicoHarp 300, PicoQuant) to analyze the time-resolved coincidence-counting events, where the resolving time-bin is set to 512 ps. The output signals are also simultaneously fed to the counting electronics to record the single- and coincidence-counting events.

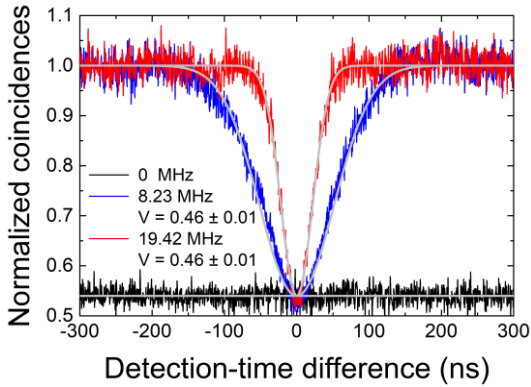

Fig. 2. Normalized coincidences as a function of the detection-time difference between the two single-photon detectors. The Hong-Ou-Mandel-type two-photon interference fringes are measured for three different frequency noises applied to the phase modulator (see text for details).

First, we performed the time-resolved measurement of the HOM fringes by employing phase-randomized CW-mode weak coherent lights. In this case, in order to determine the mutual coherence time of the two interfering photons, we utilized the PM without operating the IM. The time-resolved TPI fringes are measured for three different values of the frequency noise applied to the PM. Figure 2 shows the normalized coincidences as a function of the detection-time difference between the two SPDs D1 and D2. The light-gray solid lines represent the theoretical curve fits to the experimental results.

When there is no frequency noise in the two interfering photons, the time-delayed photons from the two interferometer arms have identical frequencies [25-27], thereby making the two-photon states at the PBS3-output indistinguishable, which therefore means that the two detected photons in the two SPDs through the PBS3 always contribute to HOM interference. The normalized coincidence is found to be $0.54 \pm 0.02$ regardless of the detection-time difference. This is the characteristic feature of the time-resolved coincidence measurement with two temporally separated photons that originate from a common light source.

On the other hand, when there is an additional frequency noise in one of the photons passing through the PM, a distinguishable frequency difference is introduced between the two time-delayed photons. Consequently, the induced "different-frequency" information determines the mutual coherence time of the two interfering photons. Here, the mutual coherence time is inversely proportional to the standard deviation of the frequency noise applied to PM, which can be identified by the FWHM of the HOM-dip fringe (Fig. 2). From the theoretical fitting, the fringe widths and visibilities for applied frequency noises are found to be $121.47 \pm 0.72$ ns and $0.46 \pm 0.01$ for 8.23 MHz and $51.49 \pm 0.44$ ns and $0.46 \pm 0.01$ for 19.42 MHz, respectively.

Next, we conducted the TPI experiment with pulse-mode weak coherent lights passed through the IM at the interferometer input. The pulse width and repetition rate were fixed at 100 ns and 2 MHz, respectively, to investigate the TPI visibility upon varying the ratio of $T_c/T_p$ in Eq. (4), because the minimum $T_c$ of ~50 ns could be obtained in our experiment. Under this condition, the mean photon-number per pulse was ~0.1. When the two pulses have orthogonal polarizations, the normalized coincidence for the detection-time difference of $\gg 100$ ns is ~0.013, which corresponds to an extinction ratio of 18.86 dB. To investigate the effect of the $T_c/T_p$ ratio on the TPI visibility, we varied the mutual coherence time by varying the frequency noise applied to the PM while the pulse duration was fixed at 100 ns. Under this condition, we measured the coincidence counts for both parallel ($\parallel$) and orthogonal ($\perp$) polarizations of the two interfering weak coherent pulses travelling through the two interferometer arms. The choice of the parallel and orthogonal polarizations in our experiment is achieved by setting the orientation angles of the H2 (with its axis oriented at 22.5° or 0°) in Fig. 1.

Figure 3 shows the experimental result of the time-resolved measurement of the TPI with weak coherent pulses. The light-gray solid lines represent the theoretical curves calculated with Eq. (2) for three different values of $T_c$. Here, the maximum visibility ($V_m$) is found to be 0.46 regardless of the coherence time, which indicates that the peak value for $t=0$ in Fig. 3 is constant. As per Eq. (3), the TPI-fringe visibility is calculated to be 0.18, 0.29, and 0.41 for 45, 85, and 250 ns of the coherence time, respectively. In the time-resolved TPI measurement of the weak coherent pulses, the loss of indistinguishability is revealed by the increase in the integrated coincidence counting rate within the pulse duration.

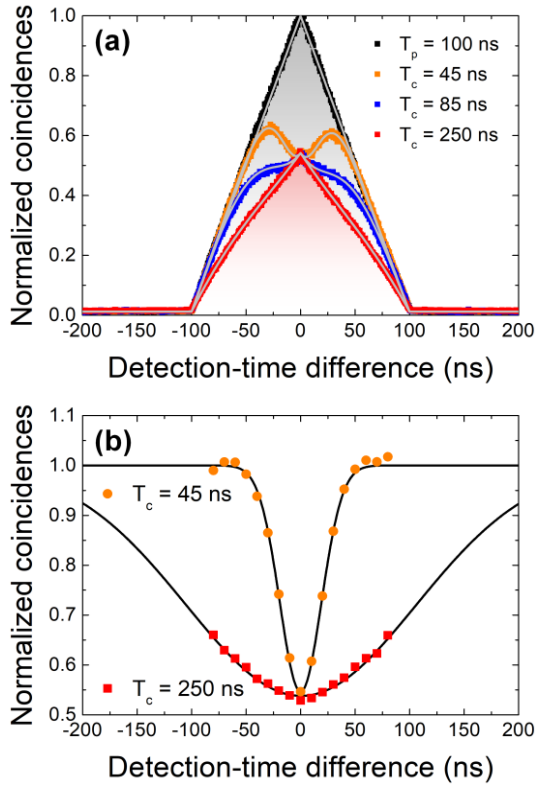

Fig. 3. Normalized coincidence as a function of the detection-time difference between the two single-photon detectors. (a) The Hong-Ou-Mandel (HOM)-type two-photon interference fringes measured by time-resolved coincidence-detection technique for three different coherence times 45 ns, 85 ns, and 250 ns. (b) The reconstructed HOM fringes for $T_c$ = 45 ns and $T_c$ = 250 ns with fixed $T_p$ = 100 ns, which are calculated from the normalized coincidences for parallel ($\parallel$) and orthogonal ($\perp$) polarizations (see text for details).

Figure 3(b) represent the conventional HOM-dip fringes for $T_c$ = 45 ns and $T_c$ = 250 ns with fixed $T_p$ = 100 ns, which are calculated from the normalized coincidences for parallel ($\parallel$) and orthogonal ($\perp$) polarizations by limiting the integration time ($T_R$) to 10 ns. Here, the HOM-dip visibility can easily be obtained from Eq. (3) by replacing $T_p$ with $T_R/2$, which corresponds to the maximum visibility for $T_R \ll T_p$. It is noteworthy that high visibility can be obtained even for a rather low $T_c$ by employing a narrow $T_R$. From the theoretical curve fitting, the fringe visibility and FWHM were found to be 0.45 ± 0.01 and 46.70 ± 0.75 ns for $T_c$ = 45 ns, and 0.46 ± 0.01 and 248 ± 3.0 ns for $T_c$ = 250 ns, respectively; these results closely agree with the counterpart values obtained by the time-resolved TPI measurement of weak coherent pulses.

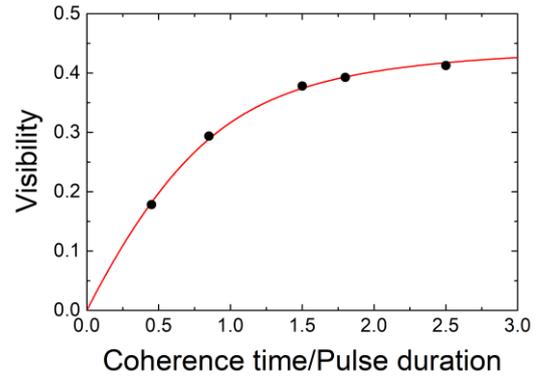

Fig. 4. Two-photon interference fringe visibility as a function of the ratio of the mutual coherence time to the pulse duration.

The TPI fringe visibility was analyzed according to the $T_c/T_p$ ratio by varying the ratio from 0.45 to 2.5, while fixing the $T_p$ at 100 ns. Figure 4 shows the observed visibility as a function of the ratio of $T_c/T_p$. The solid lines denote the theoretical curves fitted to the experimental data based on Eq. (4). The maximum visibility ($V_m$) in our experiment is found to be 0.45 ± 0.01 for the condition of $T_c \gg T_p$. When the non-zero coincidences for $t > T_p$ are subtracted, the maximum visibility improves to 0.47 ± 0.01. However, this lower visibility <0.5 may have caused by nonideal experimental conditions such as the performance of the function generator for applying the electrical signal into the PM and imperfect extinction ratio of the polarization components.

In conclusion, we have experimentally demonstrated the time-resolved TPI of the phase-randomized long-coherence weak coherent pulses and compared the result with that of the CW-mode coherent lights. The mutual coherence time was determined by adding a frequency noise to one of the two interfering weak coherent lights. The characteristic TPI-fringe shapes and visibilities were measured by varying the mutual coherence time with a fixed pulse duration. From the obtained TPI visibility according to the ratio of the coherence time to the pulse duration, it was found that the ratio pays a critical role to obtain the high visibility TPI. Our results can provide a comprehensive understanding of the time-resolved TPI fringe shape via its comparison with the conventional HOM-dip fringe. Moreover, the findings provide useful information to analyze the effect of mutual coherence between two interfering weak coherent pulses on the TPI fringe visibility.

**Acknowledgment**

This work was supported by the National Research Foundation of Korea (NRF) (2020M3E4A1080030 and 2020R1I1A1A01072979) and the Ministry of Science and ICT (MSIT), Korea, under the Information Technology Research Center (ITRC) support program (IITP-2020-0-01606) supervised by the Institute of Information & Communications Technology Planning & Evaluation (IITP). This work was also supported in part under the research program (No. 2019-106) by the affiliated institute of ETRI.

**Data availability**

The data that support the findings of this study are available from the corresponding author upon request.

**References**

1. J. L. O'Brien, A. Furusawa, and J. Vučković, "Photonic quantum technologies," Nature Photon. **3**, 687-695 (2009).
2. S. Slussarenko and G. J. Pryde, "Photonic quantum information processing: A concise review," Appl. Phys. Rev. **6**, 041303 (2019).
3. F. Flamini, N. Spagnolo, and F. Sciarrino, "Photonic quantum information processing: a review," Rep. Prog. Phys. **82**, 016001 (2019).
4. J. Wang, F. Sciarrino, A. Laing, and M. G. Thompson, "Integrated photonic quantum technologies," Nature Photon. **14**, 273-284 (2020).
5. T. F. da Silva, D. Vitoreti, G. B. Xavier, G. P. Temporão, and J. P. von der Weid, "Long-distance Bell-state analysis of fully Independent polarization weak coherent states," J. Lightwave Technol. **31**, 2881-2887 (2013).
6. H. Chen, X.-B. An, J. Wu, Z.-Q. Yin, S. Wang, W. Chen, and Z.-F. Han, "Hong-Ou-Mandel interference with two independent weak coherent states," Chin. Phys. B **25**, 020305 (2016).
7. E. Moschandreou, J. I. Garcia, B. J. Rollick, B. Qi, R. Pooser, and G. Siopsis, "Experimental study of Hong-Ou-Mandel interference using independent phase randomized weak coherent states," J. Lightwave Technol. **36**, 3752-3759 (2018).
8. Z. L. Yuan, M. Lucamarini, J. F. Dynes, B. Fröhlich, M. B. Ward, and A. J. Shields, "Interference of short optical pulses from independent gain-switched laser diodes for quantum secure communications," Phys. Rev. Applied **2**, 064006 (2014).
9. C. Agnesi, B. Da Lio, D. Cozzolino, L. Cardi, B. Ben Bakir, K. Hassan, A. Della Frera, A. Ruggeri, A. Giudice, G. Vallone, P. Villoresi, A. Tosi, K. Rottwitt, Y. Ding, and D. Bacco, "Hong-Ou-Mandel interference between independent III–V on silicon waveguide integrated lasers," Opt. Lett. **44**, 271-274 (2019).
10. H. Semenenko, P. Sibson, M. G. Thompson, and C. Erven, "Interference between independent photonic integrated devices for quantum key distribution," Opt. Lett. **44**, 275-278 (2019).
11. T. F. da Silva, D. Vitoreti, G. B. Xavier, G. C. do Amaral, G. P. Temporão, and J. P. von der Weid, "Proof-of-principle demonstration of measurement-device-independent quantum key distribution using polarization qubits," Phys. Rev. A **88**, 052303 (2013).
12. H. K. Lo, M. Curty, and B. Qi, "Measurement-device-independent quantum key distribution," Phys. Rev. Lett. **108**, 130503 (2012).
13. C. K. Hong, Z. Y. Ou, and L. Mandel, "Measurement of subpicosecond time intervals between two photons by interference," Phys. Rev. Lett. **59**, 2044-2046 (1987).
14. T. Legero, T. Wilk, A. Kuhn, and G. Rempe, "Time-resolved two-photon quantum interference," Appl. Phys. B **77**, 797-802 (2003).
15. T. Legero, T. Wilk, M. Hennrich, G. Rempe, and A. Kuhn, "Quantum beat of two single photons," Phys. Rev. Lett. **93**, 070503 (2004).
16. F. T. Arecchi, E. Gatti, and A. Sona, "Time distribution of photons from coherent and Gaussian sources," Phys. Lett. **20**, 27-29 (1966).
17. T. F. da Silva, G. C. do Amaral, D. Vitoreti, G. P. Temporão, and J. P. von der Weid, "Spectral characterization of weak coherent state sources based on two-photon interference," J. Opt. Soc. Am. B **32**, 545-549 (2015).
18. H. Kim, D. Kim, J. Park, and H. S. Moon, "Hong-Ou-Mandel interference of two independent continuous-wave coherent photons," Photon. Res. **8**, 1491-1495 (2020).
19. E. B. Flagg, A. Muller, S. V. Polyakov, A. Ling, A. Migdall, and G. S. Solomon, "Interference of single photons from two separate semiconductor quantum dots," Phys. Rev. Lett. **104**, 137401 (2010).
20. H. Bernien, L. Childress, L. Robledo, M. Markham, D. Twitchen, and R. Hanson, "Two-photon quantum interference from separate nitrogen vacancy centers in diamond," Phys. Rev. Lett. **108**, 043604 (2012).

21. P. Lombardi, M. Colautti, R. Duquennoy, G. Murtaza, P. Majumder, and C. Toninelli, "Indistinguishable photons on demand from an organic dye molecule," arXiv:2102.13055 (2021).

22. Y.-H. Deng, H. Wang, X. Ding, Z.-C. Duan, J. Qin, M.-C. Chen, Y. He, Y.-M. He, J.-P. Li, Y.-H. Li, L.-C. Peng, E. S. Matekole, T. Byrnes, C. Schneider, M. Kamp, D.-W. Wang, J. P. Dowling, S. Höfling, C.-Y. Lu, M. O. Scully, and J.-W. Pan, "Quantum interference between light sources separated by 150 million kilometers," Phys. Rev. Lett. **123**, 080401 (2019).

23. Z. Y. Ou, E. C. Gage, B. E. Magill, and L. Mandel, "Fourth-order interference technique for determining the coherence time of a light beam," J. Opt. Soc. Am. B **6**, 100-103 (1989).

24. J. G. Rarity, P. R. Tapster, and R. Loudon, "Non-classical interference between independent sources," J. Opt. B: Quantum Semiclass. Opt. **7**, S171-S175 (2005).

25. H. Kim, S. M. Lee, and H. S. Moon, "Two-photon interference of temporally separated photons," Sci. Rep. **6**, 34805 (2016).

26. H. Kim, S. M. Lee, O. Kwon, and H. S. Moon, "Observation of two-photon interference effect with a single non-photon-number resolving detector," Opt. Lett. **42**, 2443–2446 (2017).

27. D. Kim, J. Park, T. Jeong, H. Kim, and H. S. Moon, "Two-photon interference between continuous-wave coherent photons temporally separated by a day," Photon. Res. **8**, 338-342 (2020).